%% file: main.tex
\documentclass[reprint,tikz,prl,twocolumn,superscriptaddress]{revtex4-1}
\include{header}

\begin{document}

\title{Magnetic field induced intermediate quantum spin-liquid with a spinon Fermi surface}

\author{Niravkumar D. Patel}
\author{Nandini Trivedi}
\affiliation{Department of Physics, The Ohio State University, 191 W. Woodruff Avenue, Columbus, OH 43210}

\begin{abstract}
The Kitaev model with an applied magnetic field in the $H||[111]$ direction shows two transitions: from a non-abelian gapped quantum spin liquid (QSL) to a gapless QSL at $H_{c1} \simeq  0.2K$ and a second transition at a higher field $H_{c2} \simeq 0.35K$ to a gapped partially polarized phase, where $K$ is the strength of the Kitaev exchange interaction. We identify the intermediate phase to be a gapless U(1) QSL, and determine the spin structure function $S({\bf k})$ and the Fermi surface $\epsilon_F^S({\bf k})$ of the gapless spinons using the density matrix renormalization group (DMRG) method for large honeycomb clusters. Further calculations of static spin-spin correlations, magnetization, spin susceptibility, and finite temperature specific heat and entropy, corroborate the gapped and gapless nature of the different field-dependent phases. In the intermediate phase, the spin-spin correlations decay as a power law with distance, indicative of a gapless phase. 
\end{abstract}

\maketitle

\noindent 
\noindent
The inspired exact solution of the Kitaev model on a two-dimensional 
honeycomb lattice with bond dependent spin exchange interaction \cite{Kitaev1} (Eq.~\ref{eq:model}) has emerged as a paradigmatic model for quantum spin liquids (QSL). For equal bond interactions ($K$), the ground state of the Kitaev model is known to be a topologically non-trivial gapless QSL with fractionalized excitations. The multiple ground state degeneracy reflects the topology of the lattice manifold (2-fold degeneracy on a cylinder and 4-fold degeneracy on a torus). The promise of utilizing fractionalized excitations for applications in robust quantum computing \cite{Kitaev1,Blents1,ZhouRMP,knollethesis,knolle2} has led to considerable excitement and activity in the field from diverse directions, all the way from fundamental developments to applications. Upon breaking time-reversal symmetry (TRS), for example by applying a magnetic field,
the exact solvability of the Kitaev model is lost but perturbatively it can be shown that the Kitaev gapless QSL becomes a gapped QSL phase with non-abelian anyonic excitations~\cite{Kitaev1}.

In order to realize the exotic properties of the Kitaev model, there has been keen interest to discover Kitaev physics in materials. To this end, Mott insulators with magnetic frustration and large spin-orbit coupling have been proposed \cite{Khaliullin1}, leading to the discovery of $\alpha$-RuCl$_3$ and A$_2$IrO$_3$ (A = Na, Li). These are candidate QSL materials with the desired honeycomb geometry. Although given additional interactions beyond the Kitaev interaction in materials, there is still controversy about the regimes where Kitaev physics can be accessed
\cite{chun2015,Feng1,Taylor1,Banerjee2017,Winter2017,Simon2017}. 
Additionally, triangular and Kagome lattice with spin frustration has 
also been proposed as a candidate for a QSL
\cite{Misguich1,Imada2002,Motrunich2005,Saito1,YSLee1,Subir2,PLee1}.
In this context, organics such as $\kappa$-(BEDT-TTF)$_2$Cu$_2$(CN)$_3$ \cite{Shimizu1} and  EtMe$_3$Sb[Pd(dmit)$_2$]$_2$ \cite{Itou1}, the transition metal dichalcogenides  {1T-TaS$_2$}~\cite{PLee1} and a large family of rare-earth dichalcogenides AReX$_2$  (A = alkali or monovalent ions, Re = rare earth, X = O, S, Se) \cite{LiuChineseLett} have been explored. 
Remarkably, recent NMR and thermal Hall conductivity experiments on $\alpha$-RuCl$_3$ demonstrate that one can drive the magnetically ordered phase into a QSL phase using an external magnetic field~\cite{Matsuda1,Martin1}. To this end, various numerical tools were used to explore the possible QSL phases in the models relevant to candidate honeycomb materials with an external magnetic field~\cite{Wolter1,BNormand1,Gordon1,Hongchen1,takikawa1}. 

\begin{figure}[t]
\begin{center}
 \begin{overpic}[trim = -0cm 0cm 0mm 0mm,width=0.48\textwidth,angle=0]{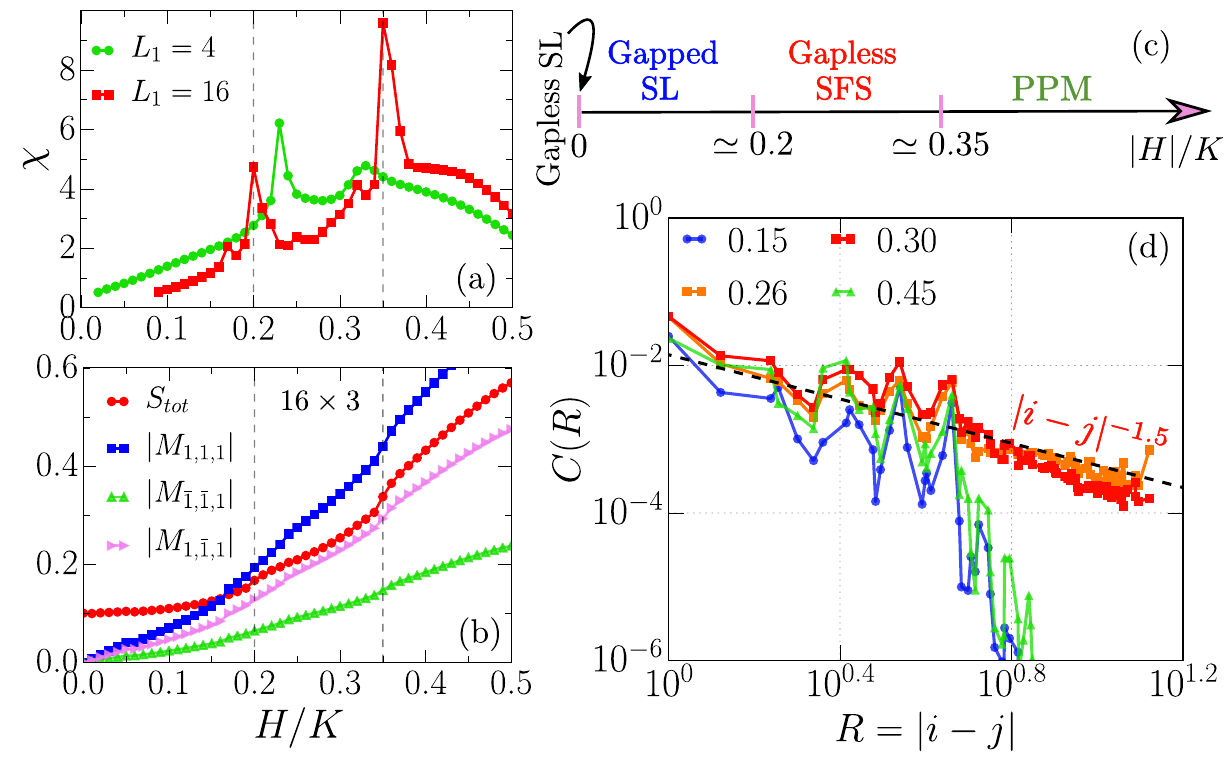}
\end{overpic}
\end{center}
\subfloat{\label{fig:1a}}
\subfloat{\label{fig:1b}}
\subfloat{\label{fig:1c}}
\subfloat{\label{fig:1d}}
\vspace{-1.0cm}
\caption{
(a) Spin susceptibility ($\chi$) for $L_1 = 4, 16$ and (b) different measures of magnetization as a function of magnetic field strength ($H$) for $L_1 = 16$. The vertical dashed lines show critical magnetic fields where phase transitions occur. (c) Corresponding phase diagram that shows transitions from gapless QSL to gapped QSL to an intermediate gapless QSL with a spinon Fermi surface (SFS) into a partial polarized magnetic (PPM) phase
($H=0$ to large $H$ respectively). (d) Average spin-spin correlations between sites at a fixed distance of $R$. The dashed black line represents the power law decay with power $-1.5$, as a guide to the eye. All Results obtained using DMRG at zero temperature.
}
\label{fig:1}
\end{figure}

In this article, we theoretically address the question: What is the fate of the Kitaev QSL with increasing magnetic field (see Eq.~\ref{eq:model}), 
beyond the perturbative limit? 
Previous studies, using a variety of numerical methods  \cite{hickey2018gapless,Ronquillo2018,Fliang1,Gohlke2018}, have pushed the Kitaev model solution to larger magnetic fields outside the perturbative regime. 
At high magnetic fields, one would expect a polarized phase. What is surprising is the discovery of an intermediate phase sandwiched between the gapped QSL at low fields and the polarized phase at high fields when a uniform magnetic field is applied along the $[111]$ direction (Fig.~\ref{fig:1c}). 
The main question we focus on is: What is the nature of this intermediate phase? 
A study based on exact diagonalization on small clusters claims to get an intermediate QSL phase with a stable spinon Fermi surface~\cite{hickey2018gapless}.  
An independent analysis based on the central charge and symmetry arguments also deduce a spinon Fermi  surface~\cite{Yming2018field,YinHe2018}. 
In this article, we present exact calculations of the spin structure factor $S({\bf k})$ on large systems and reconstruct the spinon Fermi surface $\epsilon_F^S({\bf k})$ in the intermediate phase, details of which are distinct from other recent proposals.  These are the primary significant new results of this work.


\section{Model and Method}
	We consider the Kitaev model on a honeycomb lattice in the presence of an 
external magnetic field along the $[111]$ direction. The model Hamiltonian 
is defined as 
\begin{equation} \label{eq:model}
    H_K = K \sum_{\gamma = x,y,z} \sum_{\langle i j \rangle_{\gamma}}  S^{\gamma}_i \ S^{\gamma}_j - \sum_{i} \mathbf{H}\cdot\mathbf{S}_i
\end{equation}
where $\gamma = x,y,z$ is the nearest neighbor bond index of each site, 
and $S^\gamma$ is the $\gamma$ projection of spin-$1/2$ degrees of freedom on each site. The first term is the Kitaev model where $K = 1$~eV, and the second term represents Zeeman field with a uniform magnetic field $\mathbf{H}$ (in units of $K$) applied in the $[111]$ direction. 
This model is solved using DMRG and Lanczos on a lattice with $L_1 \times L_2$ number of unit cells (u.c.) and with $2$ sites per u.c. labeled as $A$ or $B$ (the total number of sites equals $2 L_1 L_2$ - see Fig.~\ref{fig:Sura}). 
All DMRG simulations are performed with fixed $L_1 = 16$ and  
cylindrical boundary conditions using upto $1200$ DMRG states per block, ensuring the truncation error $\leq 5 \times 10^{-6}$ for $L_2 \leq 5$ ($\leq 10$ sites in $\mathbf{a_2}$ direction) within the two-site grand canonical 
DMRG~\cite{WhiteDMRG1,WhiteDMRG2,WhiteWFT,Springerbook1,
alvarez0209,alvaez3,Schollwock1,Schollwock2}. 
Additionally, we also use finite temperature Lanczos method (FTLM) for calculations of specific heat and thermodynamic entropy at finite temperatures $T$ (in units of $K$) on small clusters~\cite{Springerbook1,Prelovsek1}. 
The supplemental material presents definitions of all observables and convergence analysis of DMRG and FTLM calculations. 

%
%

\section{Results}
We begin with the magnetic and thermodynamic properties of the model. 
Figure~\ref{fig:1} shows the spin susceptibility ($\chi = \partial S_{tot}/\partial H$), and various measures of magnetization as a function of field strength $H$. 
The clear two-peak structure of 
$\chi$ demonstrate the presence of two phase transitions at 
finite field strengths. The corresponding magnetization also shows kinks at the critical fields $H \simeq 0.2$ and $0.35$ (dashed lines in Figs.~\ref{fig:1a} and \ref{fig:1b}).
The zero field limit is known to be a QSL with a gapless energy spectrum and $4$-fold topological ground-state degeneracy that is associated with a symmetry protected topological phase~\cite{Kitaev1}. 
At finite fields $H \lesssim 0.2$, the ground-state is $2$-fold degenerate on a cylinder and the energy spectrum becomes gapped~\cite{Kitaev1,Fliang1}. 
The gapped phase is also understood to be a $p+ip$ state of the emergent fermions. For high fields $H \gtrsim 0.35$, we find a partially polarized magnetic phase (PPM) phase, where the magnetization monotonically increases toward its saturation value ($M_{111} = 1.5$) in the trivially field polarized product state. 
Remarkably, the two peak structure of $\chi$ indicates that an intermediate phase is sandwiched between the gapped Kitaev QSL and polarized phases (Fig.~\ref{fig:1c}).

What are the properties of the intermediate phase? 
We study the decay of spin-spin correlations $C(R)$ shown as a function of the distance $R$ between the two spins in (Fig.~\ref{fig:1d}). As expected, in the gapped non-abelian QSL phase ($H = 0.15$) the $C(R)$ decays exponentially, consistent with a finite gap to flux  excitations\cite{Knolle1,knollethesis,Gohlke2018,hickey2018gapless}. 
In the PPM phase, the behavior of $C(R)$ is consistent with an exponential decay due to short ranged spin-spin correlations arising from the energy cost for a spin-flip that is proportional to $H$. 
However, in the intermediate regime, the behavior of $C(R)\propto R^{-m}$ is approximate fit as a power law with $m=1.5$ shown by the dashed black line in Figure~\ref{fig:1d}. The supplementary material show more detained fitting analysis of $C(R)$, where $m$ is weakly dependent on the field in the intermediate phase.

%
\begin{figure}[t]
\begin{center}
 \begin{overpic}[trim = -0cm 0cm 0mm 0mm,
 width=0.48\textwidth,angle=0]{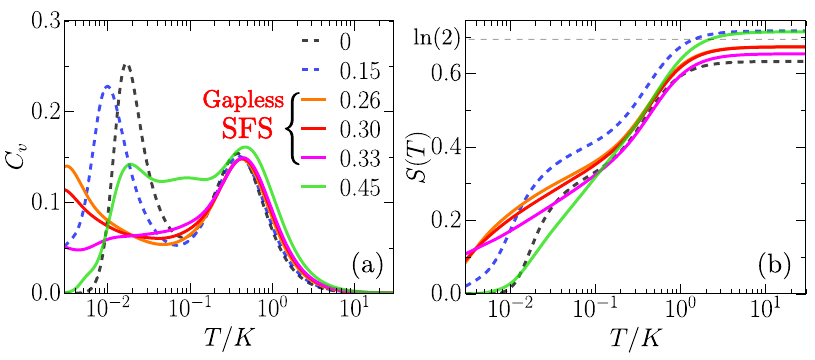}
\end{overpic}
\end{center}
\subfloat{\label{fig:2a}}
\subfloat{\label{fig:2b}}
\vspace{-1.0cm}
\caption{
Specific heat $C_v(T)$ and thermodynamic entropy $S(T)$ as a function of temperature $T$. Shown results are using a $3 \times 3$ unit cell ($18$ sites) cluster solved using finite temperature Lanczos method (FTLM) 
averaged over $50$ different different random runs~\cite{Springerbook1,Prelovsek1} for each fixed magnetic field strengths $H$ (key label).
}
\label{fig:2}
\end{figure}
%

%
\begin{figure*}[t]
\begin{center}
 \begin{overpic}[trim = 0cm 0.0cm 0mm 0mm,
width=1.0\textwidth,angle=0]{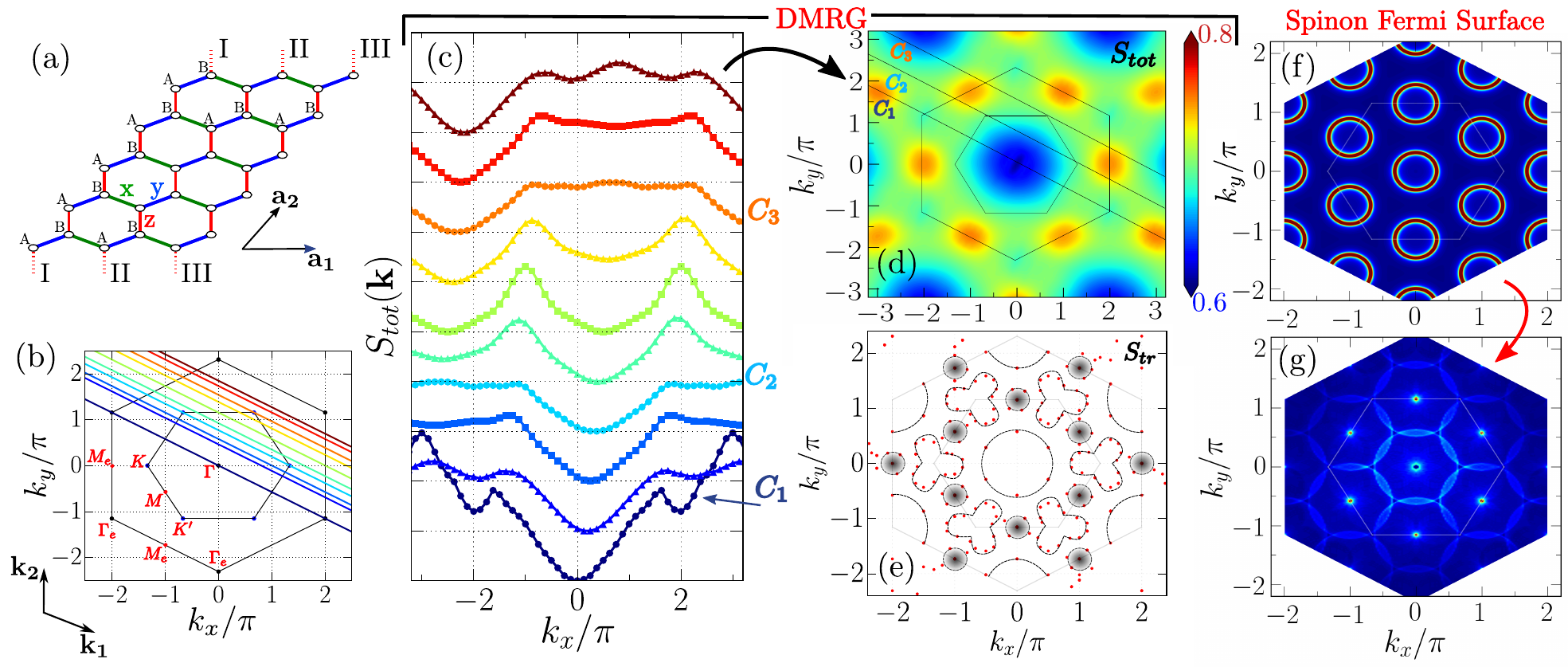}
\end{overpic}
\end{center}
\subfloat{\label{fig:Sura}}
\subfloat{\label{fig:Surb}}
\subfloat{\label{fig:Surc}}
\subfloat{\label{fig:Surd}}
\subfloat{\label{fig:Sure}}
\subfloat{\label{fig:Surf}}
\subfloat{\label{fig:Surg}}
\vspace{-1.1cm}
\caption{
The spin structure factor and deduced spinon Fermi surface in the intermediate phase. Panel (a) shows the honeycomb lattice with cylindrical boundary conditions with $L_2 = 3, 4, 5$ ($6, 8, 10$ sites in the  $\mathbf{a_2}$ direction), shown here for $L_2=5$. 
The unit vectors are $\mathbf{a_1} = (0,1)$ and $\mathbf{a_2} = (0.5,\sqrt{3}/2)$. The spin structure factor \Stotk in panel (c) shown 
along the crystal momentum cuts in panel (b), within the extended BZ, at fixed $H = 0.3$ in the intermediate phase. These cuts are defined using the reciprocal lattice vectors $\mathbf{k_1} = (2\pi,-2\pi/\sqrt{3})$ 
and $\mathbf{k_2} = (0,4\pi/\sqrt{3})$ where circles, squares, and triangle makers represent cuts obtained using $L_2 = 3, 4, 5$ respectively.  The high symmetry points of the BZ and extended BZ are also labeled in the panel (b). 
Panel (d) contour plot of \Stotk with blue representing low and red representing high intensity. The diagonal lines are especial cuts $C_1, C_2$ and $C_3$ defined in panel (c). 
(e) The peak locations (red points) of the intra-sublattice structure 
factor \Strk where grey circles at $M$ points represents large peaks. 
(f) Proposed spinon Fermi surface $A(\mathbf{k},\w=0) = \delta(\w-\epsilon^S_{F}(\mathbf{k}))$ with pockets around $\Gamma$ and $M$ points, assuming sublattice $A$ and $B$ are equivalent as in the Kitaev model. 
(g) Scattering spectrum of the proposed spinon Fermi surface calculated using 
${\cal F}({\bf {k}}) \sim \sum_{\mathbf{q}} A(\mathbf{q},\w=0) A(\mathbf{q}+\mathbf{k},\w=0)$.
The proposed spinon Fermi surface reproduce robust peaks at $M$ points of the BZ, and circular pattern around the $\Gamma$ points, similar to the DMRG results of the panel (e).  
}
\label{fig:surface}
\end{figure*}
%
We provide further evidence of the gapless nature of excitations in 
Figure~\ref{fig:2} through the temperature-dependent specific heat $C_v(T)$ and the thermodynamic entropy $S(T)$. We capture the characteristic two peak structure of $C_v$ and two `hump' structure in the entropy at zero field. As expected, $S$ saturates to ${\rm ln}(2)$ at large temperatures for all field values. 
The two peak/hump feature is also found in the gapped QSL phase ($H=0.15$), where low temperature peak/hump shifts to even lower temperatures compared to $H=0$. The behavior of the low temperature $S(T)$ and $C_v(T)$ is consistent with a gapped spectrum both at low and high fields. 
However in the intermediate phase, the peak at low $T/K$ is suppressed leading to an algebraic behavior that is distinct compared to both gapped phases. In fact, the low temperature $C_v$ at $H = 0.33$ show linear behavior in temperature (shown in supplemental material with a linear temperature scale) that is consistent with a gapless phase. 
As the FTLM is limited to small system sizes, severe size effects prevent us from obtaining an exact dependence of $C_v$ on temperature, and therefore we only discuss the qualitative trends. The analysis of the statistical error in calculations of $C_v$ and $S$ is also shown in the supplemental, along with $C_v$ at low temperatures. 
Finally with increasing $H$ across $H_{c2} \simeq 0.35$ into the PPM phase ($H = 0.45$), the two peaks in $C_v$ merge, and similarly only one hump feature is visible in the entropy. 

We also analyze the low temperature `missing' part of the entropy ($S_{low}$) resulting from finite size effects (shown in the supplemental material). The $H=0$ and gapped phases have a quadratic increase in $S_{low}$ with temperature. On the contrary,  the intermediate phase shows almost a linear behavior of $S_{low}$ at low temperatures. 
This is consistent with our earlier arguments of gapless nature of the intermediate phase. To summarize, we observe the following features in the intermediate phase: 
(1) a slow power-law decay of the $C(R)$ (Fig.\ref{fig:1d}) that is indicative of a gapless phase, and  (2) a $C_v(T)$ consistent with linear behavior at low $T$ indicating a presence of itinerant emergent fermions. 
All of this evidence taken together provide strong evidence for an intermediate phase that is indeed gapless. Additionally, recently calculated $T=0$ spin dynamics on small systems also show a continuum in energy suggesting that the intermediate phases is indeed a $U(1)$ gapless QSL~\cite{hickey2018gapless}. 
In fact, such gapless spin-liquid phase on a triangular lattice was recently shown to have stable spinon Fermi surface \cite{PLee2,Motrunich2}. Therefore, it is likely that even in our case, on a honeycomb lattice, 
we may have a spinon Fermi surface as discussed before. 

In order to explore this possibility, we calculate the spin structure factor along various cuts of the Brillouin zone (BZ) in Figures~\ref{fig:surface}. 
For two-sites per unit cell ($A$ and $B$), we define structure factor using Fourier transform of inter and intra sublattice spin-spin correlations. 
The calculation of (trace) \Strk is performed using {\it only} intra-sublattice 
correlations while (total) \Stotk sums over both inter- and intra-sublattice 
correlations (see supplemental material). 
Note that the sublattice $A$ and $B$ are located on different positions 
within a unit cell (unlike orbitals), and therefore \Stotk must be considered in the extended BZ scheme (Fig.~\ref{fig:Surb}-\ref{fig:Surd}) while the $S_{tr}(\mathbf{k})$ may be considered only using the $1^{st}$ BZ (Fig.~\ref{fig:Sure}). 
In short, $S_{tr}(\Gamma) = S_{tr}(\Gamma_e)$ , but $S_{tot}(\Gamma) \ne S_{tot}(\Gamma_e)$ where $\Gamma$ and $\Gamma_e$ refer to high symmetry points in the first and extended Brillouin zones respectively, labeled in Fig.~\ref{fig:Surb}. We shall use subscript `e' to distinguish between high symmetry points of the extended BZ versus the $1^{st}$ BZ. 

In Figure~\ref{fig:Surc}-\ref{fig:Surd}, we show cuts and contour of the \Stotk at fixed $H = 0.3K$ in the intermediate phase (see supplemental 
for cuts of the $S_{tr}(\mathbf{k}$)). There is suppression of intensity round the $\Gamma$ point and peaks around the $M_e$ points. 
The cuts between $C_2$ and $C_3$ pass through two $M_e$ points showing robust peaks visible in Fig.~\ref{fig:Surc} (green curve). 
\Stotk at all the $M_e$ points are distinctly visible as high intensity red 
circles in Fig.~\ref{fig:Surd}. Apart from the total spin structure factor $S_{tot}(\mathbf{k})$, we also analyze the peak locations of the trace spin structure factor \Strk that respects symmetries within  the $1^{st}$ BZ. Figure \ref{fig:Sure} show large peaks at the $M$ points of the $1^{st}$ 
BZ with `soft' peaks around the $K$ and $K'$ points forming $3$-petal shaped patterns, and peaks around $\Gamma$ forming a circle. In fact, comparing Figures \ref{fig:Surd} and \ref{fig:Sure}, it can be observed that the  dominant peaks at the $M$ points in \Strk simply shifts to the $M_e$ points when considering the \Stotk that also retains the inter-sublattice correlations. 

%
\begin{figure}[t]
\begin{center}
 \begin{overpic}[trim = 0.0cm 0.0cm 0mm 0mm,
 width=0.48\textwidth,angle=0]{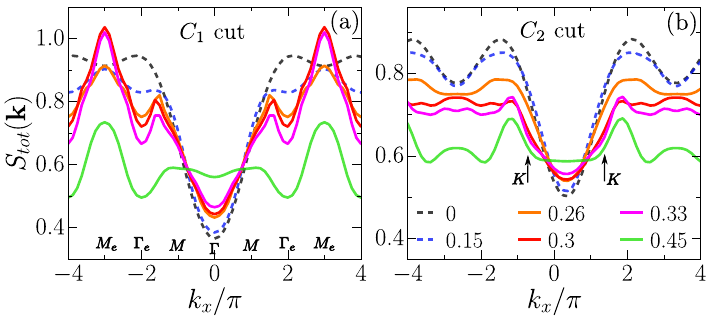}
\end{overpic}
\end{center}
\subfloat{\label{fig:SkHa}}
\subfloat{\label{fig:SkHb}}
\vspace{-1.1cm}
\caption{
\Stotk for various field strengths along (a) cut $C_1$ and (b) cut $C_2$. (defined in Fig~\ref{fig:Surd}). The peaks at $M_e$ points are most robust for the intermediate phase $H = 0.26, 0.30$ and $0.33$. The peak/hump like feature close to $K$ points are monotonically decreases with increasing $H$.
}
\label{fig:SkH}
\end{figure}
%

Given the information provided by \Stotk and \Strk, we propose the corresponding spinon Fermi surface (SFS) $\epsilon^S_{F}(\mathbf{k})$ with pockets around $\Gamma$ and $M$ points of the BZ (Fig.~\ref{fig:Surf}). 
Using the spinon spectral function, defined by $A(\mathbf{k},\w) =  \delta(\w-\epsilon^S_{F}(\mathbf{k}))$, we construct the scattering 
function in Fig.~\ref{fig:Surg}, 
${\cal F}({\bf {k}}) \sim \sum_{\mathbf{q}} A(\mathbf{q},\w=0) A(\mathbf{q}+\mathbf{k},\w=0)$, that picks up the momentum transfer ${\bf k}$ from ${\bf q}$ to ${\bf q+k}$ scattering across the Fermi surface with energy transfer $\w=0$. 
The inter-pocket scattering $\Gamma \leftrightarrow M$ and $M \leftrightarrow M$ across the FS (Fig.~\ref{fig:Surf}) leads to a large joint density of states with momentum transfer ${\bf k} = M$ that results in peaks at the $M$ points (Fig.~\ref{fig:Surg}), consistent with our DMRG calculations of \Strk in the intermediate phase (Fig.~\ref{fig:Sure}). 
Other large circular features in ${\cal F}({\bf {k}})$ arise from similar scattering processes, for example, a spinon with momentum $\mathbf{q}$ around the $\Gamma$ pocket can scatter to all $\mathbf{q+k}$ points within the pockets at the $M$ points. The combination of these large circular features leads to the 3-petal patterns also found in \Strk at the $K/K'$ points (Fig.~\ref{fig:Sure}). {\it In summary, our proposed spinon Fermi surface is able to qualitatively capture even subtle features of the \Strk in the intermediate phase providing strong evidence for a QSL with a spinon Fermi surface shown in Fig.~\ref{fig:Surf}.}

We next turn to the effect of a magnetic field on the spin structure factor. 
Figure~\ref{fig:SkH} shows \Stotk for $C_1$ cut and $C_2$ cut at various field strengths $H$. 
Note that for zero fields gapless and $H \lesssim 0.2$ gapped QSL phases, there is a broad feature that is consistent with short-ranged correlations, established by an exponential decay of spin-spin correlations in real-space (Fig.~\ref{fig:1c}). 
Our result is also consistent with previous calculations of peaks in the 
dynamical structure factor $S_{tot}(\mathbf{k},\omega)$ arising from plaquette flux and Majorana excitations \cite{Gohlke2018} that upon integrating $S_{tot}(\mathbf{k}) = \int_{0}^{\infty} d\omega S_{tot}(\mathbf{k},\omega)$, give the broad feature in Figure~\ref{fig:SkH}.

The intermediate phase with long-range correlations shows robust peaks 
at $M_e$ (Fig.~\ref{fig:SkHa}). Additionally, Figure~\ref{fig:SkHb} 
demonstrates that the plateau at momentum point $K$ monotonically decreases with an increasing magnetic field. On the other hand, $S_{tot}(\mathbf{k} = M_e)$ increase with the field, reaching its maximum value in the intermediate phase, and decreasing again upon transitioning into PPM phase. All peaks get suppressed in the PPM phase, as expected, because of shorter ranged spin-spin correlations (Fig.~\ref{fig:1d}).  
Overall, the combined results of Figures~\ref{fig:surface}-\ref{fig:SkH} show that the dominant peaks of the spin structure factor $S(\mathbf{k})$ at the $M$ points exist only in the intermediate gapless phase. 

\section{Discussion}

There have been several reports of the value of the central charge $c$ in the intermediate region, using which they have deduced the SFS~\cite{Yming2018field,Gohlke2018,YinHe2018}. However, it appears to be very difficult to obtain a reliable value for $c$ unambiguously in a gapless phase using DMRG. Ref.~\cite{Yming2018field} propose $c = 1,0$ using $L_2 = 3,4$ (respectively), Ref.~\cite{Gohlke2018} finds a $c = 4$ using $L_2 = 5$, Ref.~\cite{YinHe2018} calculate $c = 1,2$ using $L_2 = 2,3$ (respectively) using DMRG. Based on the central charge arguments, the spinon FS was argued to have pockets around the $K/K'$ and $\Gamma$ points of the first BZ~\cite{Yming2018field,YinHe2018}. This clearly differs from our proposal of the spinon FS with pockets around $\Gamma$ and $M$ points; our results are also consistent with the exact diagonalization results of the dynamical structure factor on small clusters~\cite{hickey2018gapless}. We further emphasize that the total $S(\mathbf{k})$ must respect the symmetries within the extended BZ because of two atoms in the unit cell at different locations for a honeycomb lattice (Fig.~\ref{fig:Sura}), a point that appears to have been ignored in the literature. 

We note that the $S(\mathbf{k})$ surfaces shown in Figure~\ref{fig:Surd}-\ref{fig:Sure} cannot result from tight-binding model with uniform hopping on a honeycomb lattice (the graphene problem).  This is because any scattering $K \leftrightarrow K'$ will only lead to $S(\mathbf{k})$  with peaks at the $K$ and $K'$ points, not present in our DMRG calculations of $S(\mathbf{k})$ in the intermediate phase. In order to have peaks at the $M$ point in $S(\mathbf{k})$, as we have found in our calculations, there must be pockets around the $M$ points of the spinon FS. Note that all of the $6$ $M$ points in the BZ are related by $C_3$ rotation and momentum translation symmetries, and therefore must be identical. 
Such a spinon FS with $M \leftrightarrow M$ scattering will produce a peak with momentum transfer around the $M$ point in $S(\mathbf{k})$.  Moreover, the size of pockets on the spinon FS are picked such that 
more subtle `three-petal' shaped patterns around the $K/K'$ points (Fig.~\ref{fig:Sure}) are also captured in ${\cal F}({\bf {k}})$ (Fig.~\ref{fig:Surg}). In the graphene problem, the Dirac points at the $K/K'$ are protected by inversion and time reversal symmetry. However, the magnetic field breaks time-reversal symmetry allowing for the possibility of shifting the pocket positions away from the $K/K'$ points. Additionally, we expect the largest scattering contribution to occur between a particle and hole pocket. In summary, we propose a spinon Fermi surface with two bands, creating Fermi pockets at the $\Gamma$ point and $M$ with opposite particle and hole character (Fig.~\ref{fig:Surf}).

The procedure described above for the reconstruction of the spinon Fermi  surface relies on the assumption that the $\omega-$integrated $S_{tot} (\mathbf{k}) \simeq S_{tot}(\mathbf{k},\omega=0)$ is well described by  the dynamical correlations at low energies, also used previously to propose a spinon Fermi surface in 1T-TaS$_2$~\cite{PLee2}. In the Kitaev model (at zero field) indeed most of the weight in $S_{tot}(\mathbf{k},\omega)$ is concentrated at small $\omega$~\cite{knollethesis}. We test this assumption in the new QSL phase by calculating the numerically challenging dynamical spin structure  factor~\cite{NoceraKrylov,WhiteCorrVec} at low energies $S_{tot}(\mathbf{k},\omega=0)$ on $8 \times 3$ unit cell ($48$ sites) cluster. We find that the dynamical spin structure factor also shows peaks at the $M$ points for $\omega=0$ (see supplemental), justifying our method to  reconstruct the spinon Fermi surface.

We expect the emergent neutral fermions that form a Fermi surface in the gapless QSL phase to show quantum oscillations in a magnetic field, similar to observations of quantum oscillations in SmB$_6$, a topological  Kondo insulator~\cite{SmB6_1,SmB6_2,SmB6_3}. Going forward, it would also be useful to dope the different classes of Mott insulators that could harbor quantum spin liquid ground states in order to explore the emergent superconducting phases~\cite{Subir1}.

\section{Conclusion}
Our most significant finding is that of an intermediate gapless quantum spin liquid with a spinon Fermi surface, sandwiched between the well known gapped non-abelian Kitaev spin liquid at low magnetic fields and a partially polarized phase at high magnetic fields, studied here for a field along [111] direction. 
The two quantum phase transitions are revealed by kinks in the magnetization and peaks in the susceptibility. The gapless nature of the intermediate phase $0.2 \lesssim H \lesssim 0.35$ is indicated by the slow power law decay of spin-spin correlations as opposed to an exponential decay in the other two phases. The temperature dependence of the thermodynamic quantities, specific heat and entropy, also corroborate the gapless nature of this intermediate phase.

Finally, using large-scale DMRG and detailed analysis along many cuts of  the spin structure factor $S(\mathbf{k})$ in momentum space, we propose a spinon Fermi surface in the intermediate phase with pockets at the  $\Gamma$ and $M$ points with opposite particle-hole character. 
We expect our findings to encourage the search for quantum spin liquids in materials with Kitaev interactions via spectroscopy and thermodynamic measurements.

\section{Acknowledgements} 
We would like to thank Kyungmin Lee, and 
David C. Ronquillo for their helpful comments and 
discussions. We also thank Gonzalo Alvarez for help with 
DMRG++ open source code developed at the 
Oak Ridge National Laboratory. 
N.D.P. and N.T. acknowledge support from 
DOE grant DE-FG02-07ER46423. Computations were 
performed using Unity cluster at the Ohio State University 
and the Ohio supercomputer~\cite{OSC1987}.


\bibliography{citations}

\clearpage
\pagebreak
\begin{widetext}
\clearpage
\begin{center}
\textbf{\large Supplemental: Magnetic field induced intermediate quantum spin-liquid with a spinon Fermi surface}
\end{center}
\end{widetext}

\setcounter{equation}{0}
\setcounter{figure}{0}
\setcounter{table}{0}
\setcounter{page}{1}
\makeatletter
\renewcommand{\theequation}{S\arabic{equation}}
\renewcommand{\thefigure}{S\arabic{figure}}

\section{Operators and Observables}

%
\begin{figure*}
\begin{center}
 \begin{overpic}[trim = 0cm 0.0cm 0mm 0mm,
width=0.98\textwidth,angle=0]{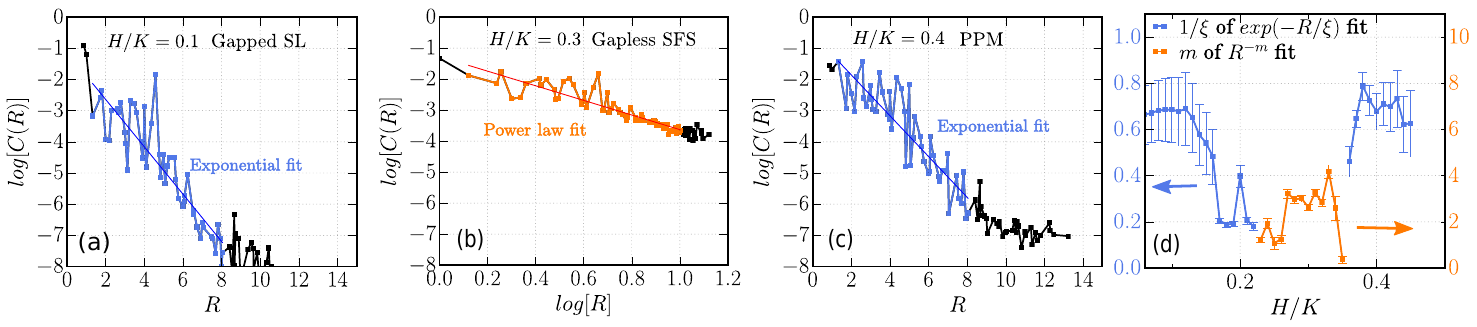}
\end{overpic}
\end{center}
\subfloat{\label{fig:SPowerVsExp_a}}
\subfloat{\label{fig:SPowerVsExp_b}}
\subfloat{\label{fig:SPowerVsExp_c}}
\subfloat{\label{fig:SPowerVsExp_d}}
\vspace{-0.7cm}
\caption{Real-space decay of spin-spin correlations fitted using an 
exponential fit ($C(R) = a \ e^{-R/\xi}$) and power law fit ($C(R) = a \ R^{-m}$) where 
$a$, $m$ and $1/\xi$ are the fitting parameters. In the gapped ($H=0.1$ - panel a) phase and 
partially polarized magnetic phase ($H=0.4$ - panel c), we use exponential decay fit (blue) 
in order to extract the decay parameter $1/\xi$ where $\xi$ represents the 
correlation length. In the gapless intermediate phase ($H = 0.3$ - panel b), we perform a fit  
using power law decay (orange) of $C(R)$. The collected decay parameters are also shown in 
panel d.
}
\label{fig:SPowerVsExp}
\end{figure*}
%

In this section, we define all the observables that are used in the main text. 
Figure~\ref{fig:1} contains results of spin susceptibility, various measures 
of magnetization, and real-space decay of the spin-spin correlation. The 
magnetization can be defined in two ways: (1) measure of the total spin 
of the ground-state $S_{tot}$ or (2) using an average of local spin moments 
that becomes finite with a finite magnetic field. 

%
\begin{equation} 
\begin{split}
S^{tot} &= \sum_{i,j} \< \mathbf{S}_i \cdot \mathbf{S}_j \> \\
M_{1,1,1} &= \frac{1}{N} \sum_{i} \< S^x_i \> + \< S^y_i \> + \< S^z_i \> \\
M_{\bar{1},\bar{1},1} &= \frac{1}{N} \sum_{i} - \< S^x_i \> - \< S^y_i \> + \< S^z_i \> \\
M_{1,\bar{1},1} &= \frac{1}{N} \sum_{i} \< S^x_i \> - \< S^y_i \> + \< S^z_i \> \\
\end{split}
\end{equation}
%
\noindent
where $S^{\gamma}_i = \frac{1}{2} \sigma^{\gamma}_i$ with $\gamma = x,y,z$ and 
$N$ is the total number of sites.
Figure~\ref{fig:1d} also show real space spin-spin correlations at fixed distance $R$ that is 
defined as 
%
\begin{equation} 
\begin{split}
C(R) &= \frac{1}{N_R} \sum_{R=|i-j|} \< {{\mathbf{S}_i} \cdot {\mathbf{S}_{j}}} \> - \< \mathbf{S}_i \> \cdot \< \mathbf{S}_j \>
\end{split}
\end{equation}
%
where $N_R$ is the total number of neighbors at distance $R$ (summed over each site). 
In the main text, we also perform calculations of specific heat ($C_v$) and entropy ($S$) as a function of 
temperature $T/K$. These calculations are performed using finite temperature Lanczos method (FTLM) on a 
$3\times3$ ($18$ sites) clusters. The details of the algorithm can be found in reference~\cite{Prelovsek1,Springerbook1}. 
%
\begin{equation} 
\begin{split}
C_v &= \frac{\partial}{\partial T} E(T) \\
S(T) &= \int_{0}^{T} dT' \ \frac{C_v(T')}{T'}
\end{split}
\end{equation}
%
where $E(T)$ is the thermal average energy calculated using $450$ number of Lanczos states. 
For each parameter $H/K$ (see main text), the $E(T)$ is calculated using 50 different 
Lanczos runs starting with a random initial state. For each run, $C_v$ and $S$ are also calculated, 
and are averaged over $50$ random FTLM runs. 

Finally, the main contribution comes from calculations of $S(\mathbf{k})$ using ground state finite DMRG. 
Note that a unit cell of honeycomb lattice has $2$ sites, located at different positions, that belong 
to sublattices $A$ and $B$. Therefore, 
we have define two different $S(\mathbf{k})$: (1) total $S_{tot}(\mathbf{k})$ that is Fourier transform 
considering all sites irrespective of the sublattice index, and (2) trace $S_{tr}(\mathbf{k})$ that 
considers only intra-sublattice spin-spin correlations. These are defined as 
\begin{equation} \label{eq:STotk}
S_{tot}(\mathbf{k}) = \frac{1}{L^2} \sum_{i,j} e^{-i \mathbf{k} \cdot \mathbf{r_{ij}}} 
\big[ \langle {{\mathbf{S}_i} \cdot {\mathbf{S}_{j}}} \rangle - \< \mathbf{S}_i \> \cdot \< \mathbf{S}_j \> \big]
\end{equation}
and 
\begin{equation} \label{eq:STrk}
S_{tr}(\mathbf{k}) = \frac{1}{L^2} \sum_{\kappa} \sum_{(i,j)_\kappa} e^{-i \mathbf{k} \cdot \mathbf{r_{ij}}} 
\big[ \langle {{\mathbf{S}_i} \cdot {\mathbf{S}_{j}}} \rangle - \< \mathbf{S}_i \> \cdot \< \mathbf{S}_j \> \big]
\end{equation}
where $\kappa$ is the sublattice index limiting the sum $i$ and $j$ such that both sites belong to the sublattice $\kappa$. 
Note that $S_{tot}(\mathbf{k})$ is periodic w.r.t. an extended Brillouin zone (BZ) because there are 
two sites in the unit cell located at physically different locations. 
However, the spinon Fermi surface (SFS) must be periodic w.r.t. $1^{st}$ BZ. Therefore, 
to develop intuition w.r.t. the first BZ, we also calculate $S_{tr}(\mathbf{k})$.

We also show calculations of the dynamical spin structure factor at fixed energy 
transfer $\omega=0$. 
In general, the real-space dynamical spin-spin correlation 
between two sites $i$ and $j$ is defined as  
%
\begin{equation} \label{eq:Sijw}
S^{\gamma \gamma'}(i,j,\w) = \frac{-1}{\pi} \operatorname{Im} \big[ 
\langle \psi_0 | {\delta S}^{\gamma}_i \ 
\frac{1}{\omega - H + E_g + i \eta} \
{\delta S}^{\gamma'}_{j}  | \psi_0 \rangle
\big],
\end{equation}
%
where $|\psi_0 \>$ represents the ground-state, $\omega$ is the 
energy transfer to the system, $\eta$ is the arbitrary broadening 
parameter, $E_g$ is the ground-state energy, and $\gamma$/$\gamma'$ 
are the $x,y,z$ projections of the spin matrices. The $\delta S^{\gamma}$ is 
defined as 
%
\begin{equation} \label{eq:dsi}
{\delta S}^{\gamma}_i = S^{\gamma}_i - \<S^{\gamma}_i\>,
\end{equation}
%
where we subtract the ground-state expectation value in order to calculate 
only the dynamical spin fluctuations. These dynamical correlators are calculated 
using DMRG within the correction-vector formulation in Krylov 
space \cite{NoceraKrylov,WhiteCorrVec}. 
In general, these 
functions are Fourier transformed into the crystal momentum domain to calculate the 
momentum-energy resolved spectra that is relevant to experiments: 
%
\begin{equation} \label{eq:Skw}
\begin{split}
S_{tot}(k,\omega) &=  \frac{1}{L^2} \sum_{i,j} e^{-ik(i-j)}  \sum_{\gamma} S^{\gamma \gamma}(i,j,\omega). \\
\end{split}
\end{equation}
%
Note that within DMRG, the site $j$ is fixed to the center of the 
lattice ($d=L/2-1$) to reduce the edge effects and computational cost, and therefore the Fourier transform 
is modified accordingly. We do not restrict the sum over the site index $i$ in order to 
measure the total (as opposed to the trace) spin dynamical structure factor.

\section{Additional Results}

%
\begin{figure}[t]
\begin{center}
 \begin{overpic}[trim = 0cm 0.0cm 0mm 0mm,
width=0.48\textwidth,angle=0]{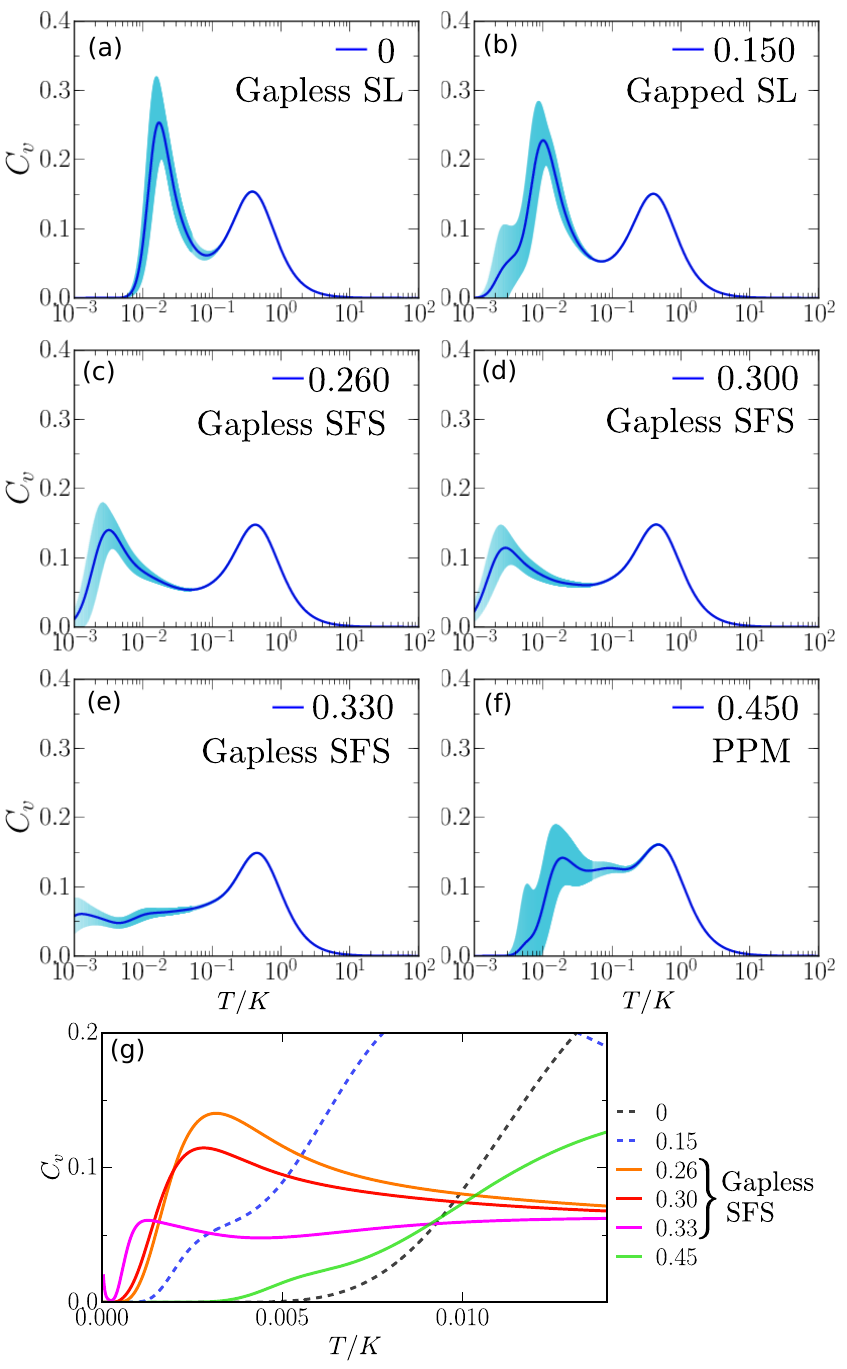}
\end{overpic}
\end{center}
\subfloat{\label{fig:Scv_a}}
\subfloat{\label{fig:Scv_b}}
\subfloat{\label{fig:Scv_c}}
\subfloat{\label{fig:Scv_d}}
\subfloat{\label{fig:Scv_e}}
\subfloat{\label{fig:Scv_f}}
\subfloat{\label{fig:Scv_g}}
\vspace{-0.7cm}
\caption{FTLM calculations of the specific heat 
as a function of temperature 
at various values of field strengths $H = 0, 0.150, 0.260, 0.300, 0.330, 0.450$. 
The blue curve is the average while shaded blue region is the error (one standard deviation) 
of $50$ independent runs. (g) The specific heat as a function of temperature in 
linear scale. In the main text, we show results $T > 0.01K$. 
}
\label{fig:Scv}
\end{figure}
%

%
\begin{figure}[t]
\begin{center}
 \begin{overpic}[trim = 0cm 0.0cm 0mm 0mm,
width=0.48\textwidth,angle=0]{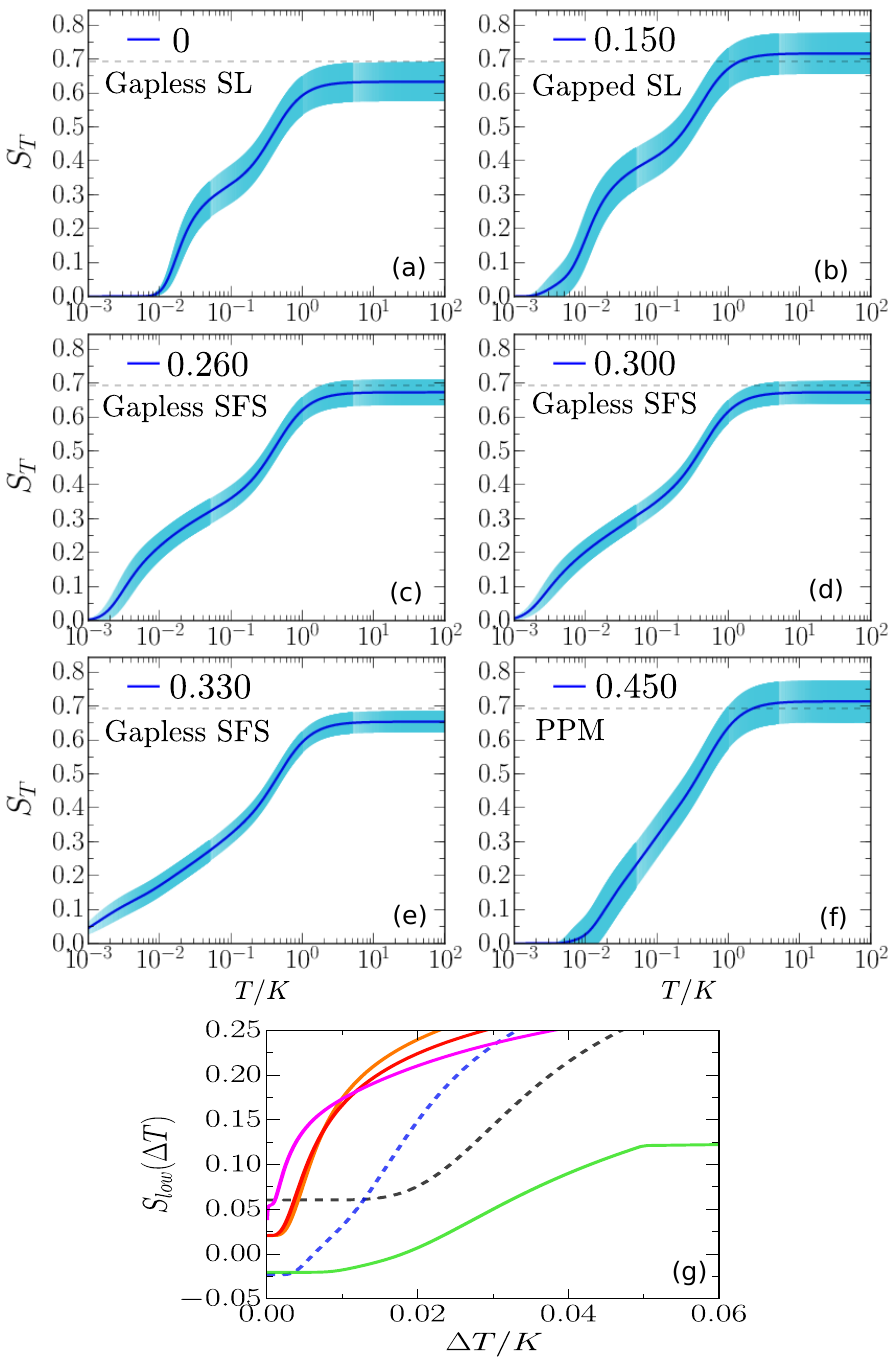}
\end{overpic}
\end{center}
\vspace{-0.7cm}
\caption{FTLM calculations of the entropy 
as a function of temperature at various values of field strengths 
$H = 0, 0.150, 0.260, 0.300, 0.330, 0.450$. 
The blue curve is the average while shaded blue region is the error (one standard deviation) 
of $50$ independent runs. The horizontal dashed grey line is $ln(2)$ representing 
large temperature entropy of spin systems. 
Panel (g) show $S_{low}(\Delta T) = ln(2) - \int^{\infty}_{\Delta T} dT' ( C_v(T')/T')$; 
residual missing part of the entropy resulting from the finite size 
effects.
}
\label{fig:SEnt}
\end{figure}
%

In this section, we show additional supporting results and various calculations performed 
to ensure the quality of the presented results. Figure~\ref{fig:SPowerVsExp} show various 
fits performed on the $C(R)$ in order to extract the decay parameter as a power in power-law 
decay or the exponent in the exponential decay. In order to perform the fitting we first 
linearize the data by using $log(C(R))$ vs. $R$ curves for the exponential fit and 
$log(C(R))$ vs. $log(R)$ for the power law fits. To have reliable fitting, 
we also disregard the $C(R)$ contributions coming 
from the boundaries, and fit the data only using the central region of $R$. 
Decay of the intermediate phase at $H = 0.3$ (Fig.~\ref{fig:SPowerVsExp_b}) 
is visibly slower than that of the gapped $H/K = 0.1$ (Fig.~\ref{fig:SPowerVsExp_a}) 
and PPM $H = 0.4$ phase (Fig.~\ref{fig:SPowerVsExp_c}). The collected 
decay parameters are explicitly plotted in Fig.~\ref{fig:SPowerVsExp_d} where 
error bars are obtained by averaging over many different central regions of 
the lattice. The blue curves are the inverse correlation length of 
an exponential fits performed for the gapped QSL ($H \lesssim 0.2K$) and 
PPM phase ($H \gtrsim 0.35$), while the 
orange curve is power $m$ of power law fits of the intermediate phase. It is clear from the  
Fig.~\ref{fig:SPowerVsExp} that decay in the intermediate phase is significantly slower than 
other phases, suggesting of a phase without a spin-gap.

In Fig.~\ref{fig:2} of the main text, we also presents results of 
specific heat $C_v$ and entropy $S$. These calculations are done using 
finite temperature Lanczos method.  
Lanczos is an iterative procedure that returns $M$ set of eigenstates of the Hamiltonian 
($M<<N_H$ where $N_H$ is the full Hilbert space) where states corresponding to the lowest and 
highest eigen-values are well converged but the intermediate states are only approximated. 
This approximated states contribute errors in calculations of finite temperature 
quantities that requires 
information about the full eigen-value spectrum. FTLM controls this error by 
averaging over many FTLM runs starting with a different random initial state ~\cite{Springerbook1,Prelovsek1}.  
In Figure~\ref{fig:Scv} and Figure~\ref{fig:SEnt}, we show 
the error (one standard deviation of $50$ independent runs) resulting 
from such averaging procedure. Note that error in $C_v$ calculation 
is almost zero at high temperatures, but the intermediate to low temperature has 
finite error shown by shaded blue region. Note that 
because we use finite $3 \times 3$ honeycomb cylindrical clusters, our results 
at low temperature are severely effected by finite size effect (finite-size gap in particular). 
In order to understand the low temperature behavior of $C_v$, we also show $C_v$ in 
linear scale at low $T/K$ (Fig.~\ref{fig:Scv_g}).  
In Fig.~\ref{fig:Scv_a}-\ref{fig:Scv_f}, we show data only for $T/K > 10^{-3}$, although  
results below $T/K \lesssim 10^{-2}$ may be significantly plagued by finite size effects and 
should be understood only as qualitative trends. 
Similar to $C_v$, we also show error in calculations of entropy in Figure~\ref{fig:SEnt}. 
However, reliability of our results is evident looking at the high temperature entropy $S$ 
correctly converging to ${\rm ln}(2)$ for all phases (as expected for spin systems). 
To further understand the error arising from finite size effects, we show 
$S_{low}$, the missing part of entropy, as a function of temperature. $S_{low}$ 
clearly show a quadratic behavior in temperature for gapped and PPM phase, while 
the gapless phase has a distinct behavior in $S_{low}$ at low temperatures resembling 
almost a linear increase with $T/K$.

%
\begin{figure}[t]
\begin{center}
 \begin{overpic}[trim = 0cm 0.0cm 0mm 0mm,
width=0.48\textwidth,angle=0]{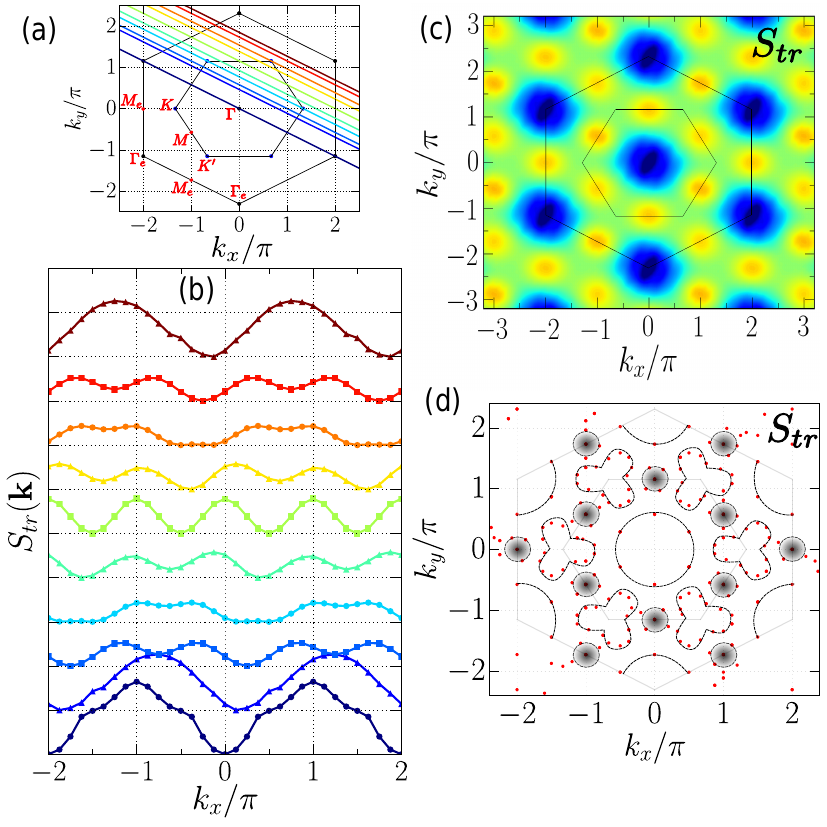}
\end{overpic}
\end{center}
\subfloat{\label{fig:SStra}}
\subfloat{\label{fig:SStrb}}
\subfloat{\label{fig:SStrc}}
\subfloat{\label{fig:SStrd}}
\vspace{-0.7cm}
\caption{The spin structure factor \Strk in panel (b) shown 
along the crystal momentum cuts in panel (a), within the extended BZ, at fixed 
$H/K = 0.3$ in the intermediate phase. 
These cuts are defined using the reciprocal lattice 
vectors $\mathbf{k_1} = (2\pi,-2\pi/\sqrt{3})$ 
and $\mathbf{k_2} = (0,4\pi/\sqrt{3})$.  The high 
symmetry points of the BZ and extended BZ are also labeled.
Panel (c) contour plot of \Strk with blue representing low and red 
representing high intensity. 
(e) The peak locations (red points) of the intra-sublattice structure 
factor \Strk where grey circles at $M$ points represents large peaks, also 
clearly visible as red circles in panel (c) contour plot. 
}
\label{fig:SStr}
\end{figure}
%

%
\begin{figure}[b]
\begin{center}
 \begin{overpic}[trim = 0cm 0.0cm 0mm 0mm,
width=0.48\textwidth,angle=0]{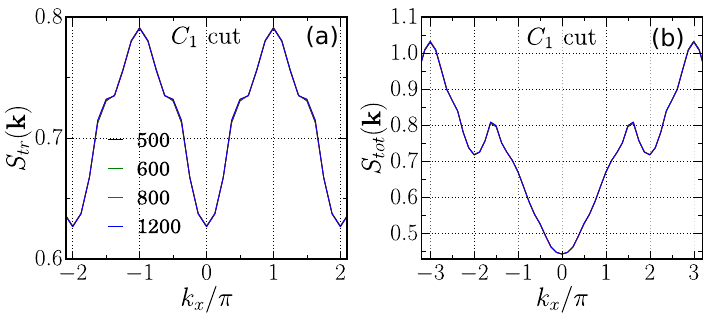}
\end{overpic}
\end{center}
\vspace{-0.7cm}
\caption{(a) $S_{tr}(\mathbf{k})$ and (b) $S_{tot}(\mathbf{k})$ 
for increasing number of states (key label) calculated using a $16 \times 3$ 
honeycomb cylinder at fixed $H/K = 0.3$. The presented curves are for the $C_1$ cut 
(defined in Fig.~\ref{fig:Surd}) of the BZ.
}
\label{fig:Smscale}
\end{figure}
%
%
\begin{figure}[t]
\begin{center}
 \begin{overpic}[trim = 0cm 0.0cm 0mm 0mm,
width=0.48\textwidth,angle=0]{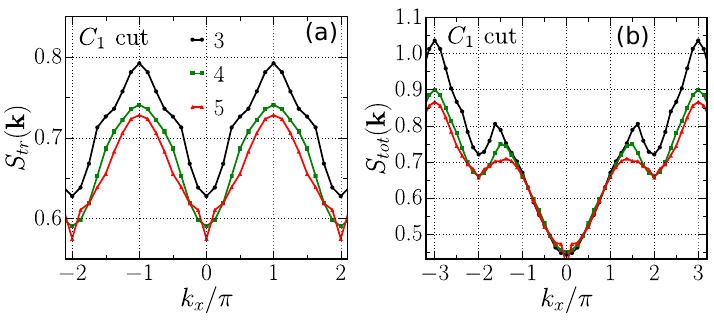}
\end{overpic}
\end{center}
\vspace{-0.7cm}
\caption{(a) $S_{tr}(\mathbf{k})$ and (b) $S_{tot}(\mathbf{k})$ 
for increasing number of sites in the $L_2$ direction (key label) with a cylindrical 
geometry using up to $1200$ states at fixed $H/K = 0.3$. 
The presented curves are for the $C_1$ cut 
(defined in Fig.~\ref{fig:Surd}) of the BZ.
}
\label{fig:SL2scale}
\end{figure}
%

%
\begin{figure}[b]
\begin{center}
 \begin{overpic}[trim = 0cm 0.0cm 0mm 0mm,
width=0.48\textwidth,angle=0]{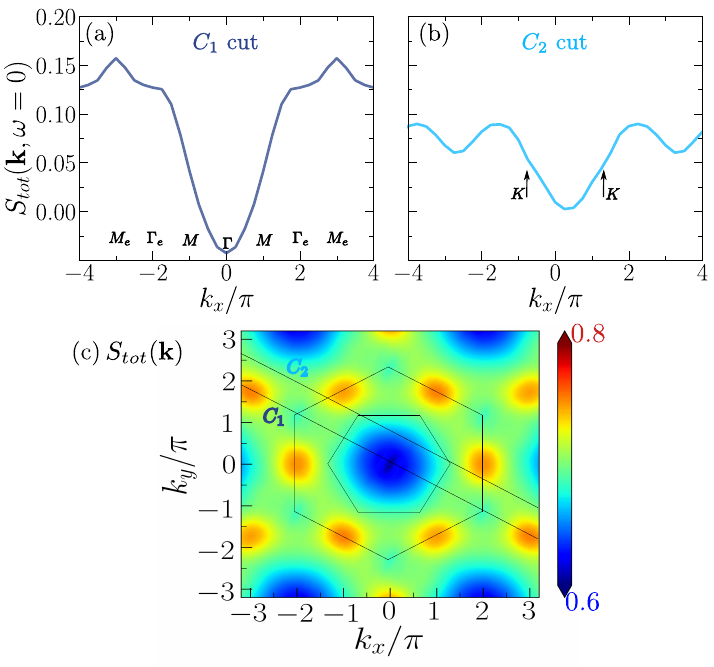}
\end{overpic}
\end{center}
\subfloat{\label{fig:SSkwa}}
\subfloat{\label{fig:SSkwb}}
\subfloat{\label{fig:SSkwc}}
\vspace{-0.7cm}
\caption{The zero energy transfer 
dynamical spin response function $S_{tot}(\mathbf{k},\w=0)$ 
for the cuts (a) $C_1$ and (b) $C_2$ of within the BZ. This quantity represents 
scattering of the emergent fermions across the FS with zero energy transfer. (c) 
The total static spin structure factor $S_{tot}(\mathbf{k})$ where the cuts 
$C_1$ and $C_2$ are represented by the diagonal black lines that cuts through the 
$1^{st}$ and $2^{nd}$ BZ. 
}
\label{fig:SSkw}
\end{figure}
%

For completeness, Figure~\ref{fig:SStr} show cuts of the \Strk, 
the peak locations of the \Strk and the 
corresponding contour plot. The lowest dark-blue cut 
shown in Fig.~\ref{fig:SStrb} show robust peaks at 
the $M$ points ($k_x = \pi$), and in turn these peaks 
at $M$ points are also visible in the contour plot 
as red circles and the grey circles in the 
plot of peak locations (Fig.~\ref{fig:SStrc}-\ref{fig:SStrd}). 
The 3-petal shaped patterns around the 
$K/K'$ points are not clearly visible in the contour plot as they are 
over-shadowed by the large peaks at the $3$ neighboring $M$ points.

To check the quality of presented data, we also show scaling of the 
presented observable in the main text. We first show scaling 
of the spin structure factor with increasing 
number of states using $16\times3$ cylinders (Figure~\ref{fig:Smscale}). 
The results do not change with increasing the number of DMRG states, and therefore 
we demonstrate that results of $S(\mathbf{k})$ are well converged 
even using only $500$ DMRG states. Finally, 
Figure~\ref{fig:SL2scale} show $S(\mathbf{k})$ for 
increasing lattice size in the $L_2$ direction while 
using fixed $L_1 = 16$ and $H/K = 0.3$. It is not 
surprising that $S(\mathbf{k})$ peaks are renormalized with increasing 
system size. 
However, we show that peaks at the $M$ ($k_x/\pi = \pm 1$) and $M_e$ ($k_x/\pi = \pm 3$) points of the $1^{st}$ and extended BZ
remain robust in all systems. Therefore, we expect that these dominant peaks at $M$ points will remain a $2D$ 
extension of the Honeycomb lattice used within DMRG.

In the main text, we used the $S_{tot}(\mathbf{k})$ and $S_{tr}(\mathbf{k})$ in order to 
reconstruct the Fermi surface using the scattering arguments across the Fermi surface. 
Scattering arguments on the FS imply that there is no energy transfer to the 
system, as these scatterings occur at a fixed energy, and therefore one should calculate the 
$S(\mathbf{k},\w=0)$ to reconstruct a FS. However, 
we have made an assumption that peaks in the static 
structure factor `mostly' result from the low energy scattering across the spinon FS, 
i.e. $S(\mathbf{k}) \simeq S(\mathbf{k},\w=0)$. Figure~\ref{fig:SSkw} show that indeed 
this is not a bad assumption. Figure~\ref{fig:SSkwa}-\ref{fig:SSkwb} show $S_{tot}(\mathbf{k},\w=0)$ 
cuts $C_1$ and $C_2$. Comparing the results of $S_{tot}(\mathbf{k}) \simeq S(\mathbf{k},\w=0)$ 
(Fig~\ref{fig:SSkwa}-\ref{fig:SSkwb}) with the static $S_{tot}(\mathbf{k})$ (Fig~\ref{fig:SSkwc}) clearly 
show that the high intensity peaks (red circles) of $S_{tot}(\mathbf{k})$
match almost exactly with the peaks found the dynamical response. These results further supports 
our conjectured Fermi surface that is show in the main text.


\end{document}

%% file: header.tex
\AtBeginDocument{
\heavyrulewidth=.08em
\lightrulewidth=.05em
\cmidrulewidth=.03em
\belowrulesep=.65ex
\belowbottomsep=0pt
\aboverulesep=.4ex
\abovetopsep=0pt
\cmidrulesep=\doublerulesep
\cmidrulekern=.5em
\defaultaddspace=.5em
}

\usepackage[usenames,dvipsnames]{xcolor}
\usepackage{amsmath,amssymb,graphicx,float,tikz}
\usepackage{amsthm,comment,booktabs}
\usepackage[hidelinks]{hyperref} 
\usepackage[printwatermark]{xwatermark}
\usepackage[export]{adjustbox}
\usepackage[section]{placeins}
\usepackage[caption=false]{subfig}
\usepackage[percent]{overpic}

\definecolor{red1}{HTML}{FF4136}
\definecolor{green1}{HTML}{00802b}

\hypersetup{colorlinks=true,citecolor=Red,linkcolor=Blue,urlcolor=Blue}
\usepackage[export]{adjustbox}

\newcommand{\<}{\langle}
\renewcommand{\>}{\rangle}
\newcommand{\w}{\omega}

\newcommand{\Stotk}{$S_{tot}(\mathbf{k})$ }
\newcommand{\Strk}{$S_{tr}(\mathbf{k})$ }